\documentclass[conference,10pt]{IEEEtran}
\usepackage{epsfig,epsf,rotating,setspace,latexsym,amsmath,amssymb,amsfonts,bm,theorem,subfigure,epstopdf}
\usepackage{cite,authblk}
\usepackage{bbm}
\usepackage{algorithm}
\usepackage[noend]{algpseudocode}
\usepackage{color}
\usepackage{mathtools}
\usepackage{soul}

\algrenewcommand\algorithmicforall{\textbf{foreach}}
\algrenewcommand\algorithmicindent{.8em}

\IEEEoverridecommandlockouts
\allowdisplaybreaks

\begin{document}
 
\title{Information Mutation and Spread of Misinformation in Timely Gossip Networks}
 
\author{Priyanka Kaswan \qquad Sennur Ulukus\\
        \normalsize Department of Electrical and Computer Engineering\\
        \normalsize University of Maryland, College Park, MD 20742\\
        \normalsize  \emph{pkaswan@umd.edu} \qquad \emph{ulukus@umd.edu}}
 
\maketitle

\begin{abstract}
We consider a network of $n$ user nodes that receives updates from a source and employs an age-based gossip protocol for faster dissemination of version updates to all nodes. When a node forwards its packet to another node, the packet information gets mutated with probability $p$ during transmission, creating misinformation. The receiver node does not know whether an incoming packet information is different from the packet information originally at the sender node. We assume that truth prevails over misinformation, and therefore, when a receiver encounters both accurate information and misinformation corresponding to the same version, the accurate information gets chosen for storage at the node. We study the expected fraction of nodes with correct information in the network and version age at the nodes in this setting using stochastic hybrid systems (SHS) modelling and study their properties. We observe that very high or very low gossiping rates help curb misinformation, and misinformation spread is higher with moderate gossiping rates. We support our theoretical findings with simulation results which shed further light on the behavior of above quantities.
\end{abstract}

\section{Introduction}\label{sec:intro}

\begin{figure}[t]
\centerline{\includegraphics[width=0.75\columnwidth]{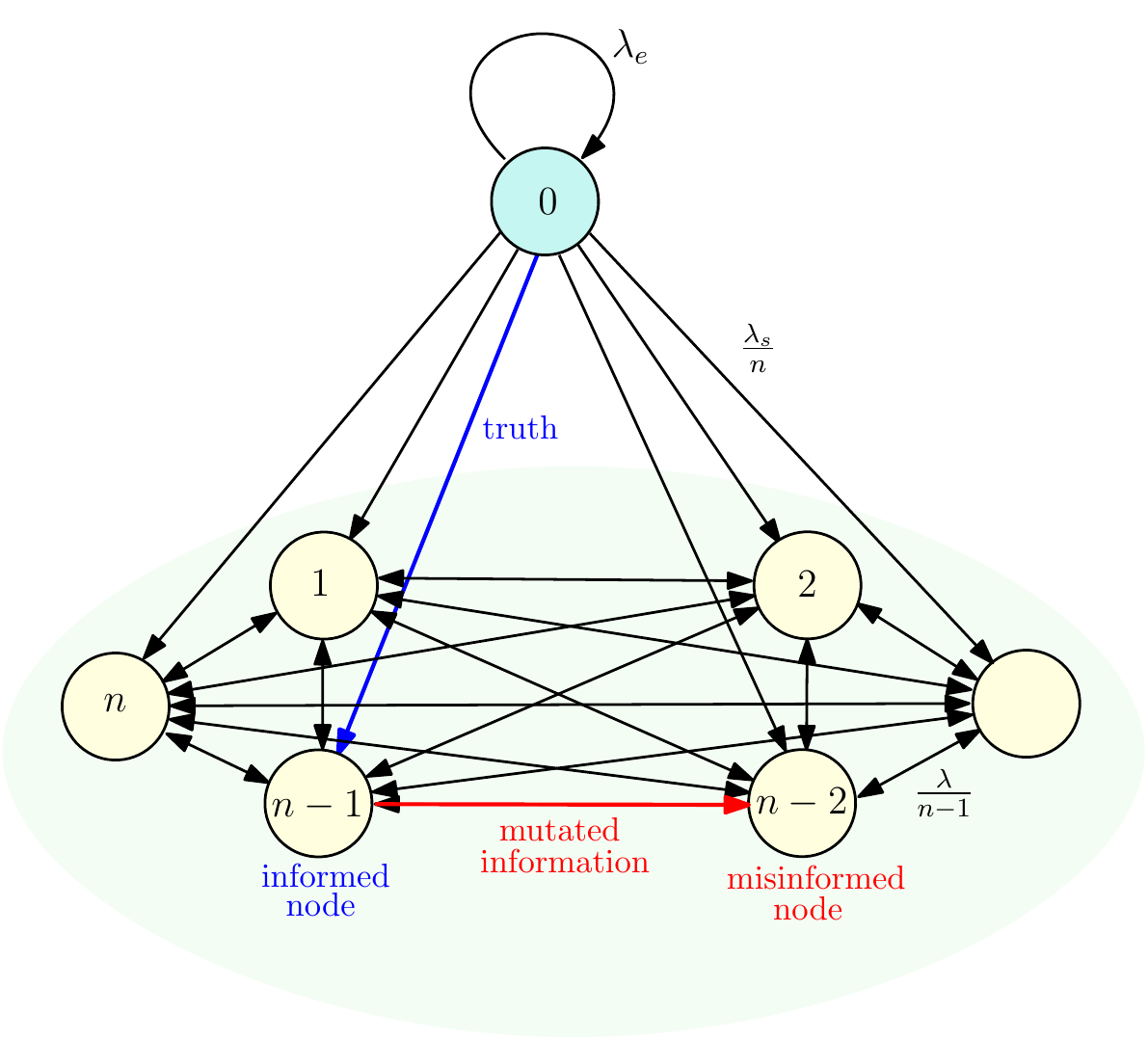}}
\caption{Fully connected gossip network of $n$ nodes with information probabilistically mutating into misinformation during internode gossip.}
\label{fig:system_model}
\vspace*{-0.4cm}
\end{figure}

In this work, we attempt to characterize spread of misinformation in an age-based gossip network, where information at the source node gets updated according to a Poisson process with rate $\lambda_e$; Fig.~\ref{fig:system_model}. We associate a version number with each information generated at the source, such that the version number of the current information at the source gets incremented by one post each source update. The source forwards its latest version to the network of $n$ nodes, $\mathcal{N}=\{1,\ldots,n\}$ according to a Poisson process with rate $\lambda_s$, choosing the destination node uniformly at random from $\mathcal{N}$ each time. The network nodes wish to have access to the latest possible version of information, and therefore, each node only stores the latest version of information it has received so far and gets rid of all older information packets. Further, the nodes gossip with their neighboring nodes to further improve the timeliness of information in the network, whereby each node sends updates to its neighbors according to a Poisson process with rate $\lambda$. 

The source always communicates accurate information to network nodes, however, there is a possibility of information getting mutated during inter-node transmissions in the network. This could either be because the sender node is not always honest and sometimes deliberately tampers with the information before forwarding it, or the information packet could get corrupted during transmission process. An example for this problem setting could be software distribution to end users, where the software vendor always provides reliable versions or iterations of software packages to users, however if the users instead obtain a software version from their neighboring users, they might occasionally receive an incorrectly functioning or even harmful version of the software. Other examples could be real time news dissemination in a region with truth getting mutated in inter-personal gossip, or smart sensor networks where occasionally noisy measurements are transmitted.

Network nodes want information that is both fresh and accurate. We use version age of information metric to characterize the freshness of information at nodes. If $V_s(t)$ corresponds to version number of the packet prevailing at the source at time $t$ and $V_i(t)$ corresponds to the version number marked on the packet present at node $i$, then the instantaneous version age of information at node $i$ at time $t$ is defined as $X_i(t)= V_s(t)-V_i(t)$. Further, we use $T_i(t)$ to denote the accuracy of information at node $i$, with $T_i(t)=1$ implying node $i$ has accurate information (alternatively referred to as the truth), and $T_i(t)=0$ implying that node $i$ has inaccurate information (alternatively referred to as misinformation). 

Whenever node $i$ receives a packet, it compares the version number of the received packet with the version number of the packet in its possession, and if the version numbers are different, then node $i$ discards the staler packet and keeps the fresher packet. However, node $i$ does not prima facie know whether the piece of information it has received is the truth or not, or if it is different from sender's information or not. Therefore, if the information already present at node $i$ and the received information have the same version numbers, node $i$ just keeps the information it trusts the most, say based on performance of software in the software distribution example, or measurement noise detected in smart sensor network example. If one of the packets contains accurate information (i.e., the truth), then node $i$ would keep that packet, i.e., truth prevails over misinformation in our model. In this regard, we wish to know what fraction of nodes are misinformed in this network.

Gossip algorithms are decentralized algorithms which involve network nodes randomly contacting their neighbors to exchange packets, and traditionally, their analysis is done using the epidemic model of data spread \cite{bailey75} where the quantity of interest is dissemination time of a fixed message or set of messages to all nodes of the network\cite{Demers1987EpidemicAF-short, Minsky02cornellthesis, vocking2000, Pittel1987OnSA, deb2006AlgebraicGossip, devavrat2006, Sanghavi2007GossipFileSplit}. Likewise, misinformation literature \cite{zhao13, Nekovee07, Friggeri14, Moreno04, Chierichetti09, Acemoglu10} also treats misinformation as a virus, such that users can become infected upon exposure, consequently turning them into spreaders. This allows for interpretation of misinformation spread as epidemic models like SIS and SIR, and the quantity of interest here is how long misinformation survives in a network.  Though both gossip and misinformation spread are essentially information diffusion processes, the goals in the two cases are different, since rapid and complete spread of an update is preferred in the former. 

The above works consider dissemination of static information in the network. However, with the highly dynamic nature of data sources in modern applications, network nodes are interested in the most up-to-date information at all times, which has recently motivated works in timely or age-based gossiping \cite{Yates21gossip, baturalp21comm_struc, kaswan22slicingcoding, Mitra23opportunisitic,elmagid23gossipagedist, bastopcu_agent_gossip, kaswan22timestomping, Kaswan23reliable}. Since our paper considers the subject of misinformation in timely gossip networks, the works that are most closely related to our paper are \cite{bastopcu_agent_gossip, kaswan22timestomping, Kaswan23reliable}: \cite{bastopcu_agent_gossip} studies the conditions under which the majority rule based estimation of time-varying binary valued source can cause incorrect source estimation or misinformation at network nodes.  In \cite{kaswan22timestomping}, an adversary alters timestamps of packets, rebranding old packets as fresh and fresh packets as old, with the goal to replace circulation of fresh packets with outdated packets in the network. \cite{Kaswan23reliable} considers network model with two sources, such that the information obtained from one source is considered more reliable than the other source, and the network nodes prefer a reliable packet over an unreliable packet even when the former is a bit outdated with respect to the latter. 

Our goal is somewhat similar to \cite{Kaswan23reliable}, as we are interested in finding what fraction of user nodes on average in the network have the truth and what is the version age of information at user nodes. However, in \cite{Kaswan23reliable}, once a packet is created at one of the sources, the packet information does not change during the packet diffusion process, whereas in the current paper, information is susceptible to mutating into misinformation during inter-node transmissions. Further, in \cite{Kaswan23reliable}, network nodes know if a particular packet has originated at the reliable source or the unreliable source, which allows nodes to consider a freshness-reliability trade-off. However, in the current work, nodes do not know whether a received information is the truth or not. In this respect, we model the information mutation problem as a stochastic hybrid system (SHS), and study the dependency of our results on various network parameters. We observe that, while very low gossip rates control the dissemination of mutated information on one hand, very high gossip rates help disseminate accurate fresh packets to all network nodes faster on the other hand. Thus, both extreme cases help curb misinformation and misinformation spread is higher with moderate gossip rates.

\section{System Model and SHS Characterization}\label{sec:sysmodel}

The system model consists of source node (node $0$), that gets version updates with rate $\lambda_e$, and $n$ user nodes $\mathcal{N}=\{1,\ldots,n\}$ that wish to have access to accurate and latest possible version of information. The source sends version updates to node $i \in \mathcal{N}$ on $(0,i)$ link according to a thinned Poisson process with rate $\frac{\lambda_s}{n}$. For $i,j\in \mathcal{N}$, node $i$ sends updates to node $j$ according to a thinned Poisson process with rate $\frac{\lambda}{n-1}$. The source possesses accurate and latest information at all times, i.e., $T_0(t)=1$ and $X_0(t)=0$, for all $t$. The source always communicates the truth to the user nodes, i.e., information does not mutate on $(0,i)$ links. However, the user nodes are not always honest. When node $i$ sends a packet to node $j$ on link $(i,j)$ at time $t$, node $i$ either honestly forwards the information in its possession with probability $1-p$ or alters the contents of the packet to send misleading information with probability $p$, spreading misinformation in the latter case. When node $j$ receives the packet from node $i$ at time $t$, $T_j(t)$ and $X_j(t)$ are reset according to following state reset protocol:
\begin{itemize}
    \item If $X_i(t^-)>X_j(t^-)$, node $j$ rejects the incoming packet and $T_j(t),X_j(t)$ pair remains unchanged, since node $j$ already has the fresher packet.
    \item If $X_i(t^-)<X_j(t^-)$, the incoming packet corresponds to a fresher version, hence node $j$ replaces its packet with the incoming packet. Note that node $j$ does not know if node $i$ was honest or not, or the received packet is the truth or not. Hence $X_j(t)=X_i(t^-)$ and
\begin{align*} 
    T_j(t)=
    \begin{cases}
        T_i(t^-), & \text{node $i$ is honest} \\
        0, & \text{node $i$ is not honest}
    \end{cases}
\end{align*}    
    \item If $X_i(t^-)=X_j(t^-)$, both packets are equally fresh and $X_j(t)$ remains unchanged. Further, in our model, truth prevails over misinformation when two information packets of the same version are encountered and either of them carries the truth. The accuracy of the incoming packet is $T_i(t^-)$ if node $i$ is honest in its communication, and $0$ if node $i$ is dishonest, while the version age of incoming packet is $X_i(t^-)$ in both cases. Hence,
\begin{align*} 
    T_j(t)=
    \begin{cases}
        \max\{T_i(t^-),T_j(t^-) \}, & \text{node $i$ is honest} \\
        T_j(t^-), & \text{node $i$ is not honest}
    \end{cases}
\end{align*}
Note that when node $i$ communicates its information honestly to other nodes, it might lead to spread of misinformation if node $i$ was misinformed to begin with ($T_i(t^-)=0$), i.e., honesty in communication does not imply delivery of the truth. 
\end{itemize}

\begin{figure}[t]
\centerline{\includegraphics[width=0.8\columnwidth]{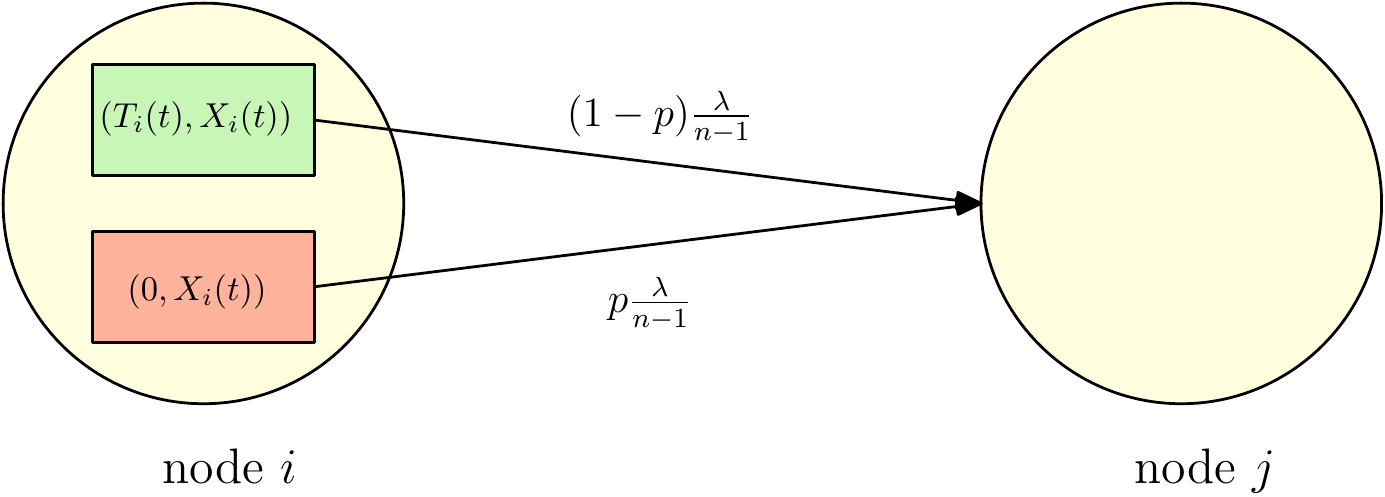}}
\caption{The green packet depict the packet actually stored at the node $i$. The orange packet depicts a mutated copy of the green packet, hypothetically available at node $i$ at all times.}
\label{fig:green_orange_packets}
\vspace*{-0.4cm}
\end{figure}

At time $t$, the fraction of users which possess the truth is
\begin{align}\label{eqn:F(t)_def}
    F(t)= \frac{T_1(t)+ T_2(t)+ \ldots + T_n(t)}{n}
\end{align}
To evaluate the expected fraction of users with truth, $F= \lim_{t \to \infty} \mathbb{E}[F(t)]$, we model the problem as an SHS. Note that node $i$ essentially sends its packet to node $j$ with honesty according to a thinned Poisson process with rate $(1-p)\frac{\lambda}{n-1}$ or sends a mutated copy of its packet according to a thinned Poisson process with rate $p\frac{\lambda}{n-1}$. For ease of exposition, we assume that instead of creating a mutated copy of its packet at the time of transmission, node $i$ stores a mutated copy of its packet at all times, i.e., each node is assumed to store two packets (see Fig.~\ref{fig:green_orange_packets}). In Fig.~\ref{fig:green_orange_packets}, the green packet represents the packet actually present at the node $i$, having accuracy of $T_i(t)$ and version age of $X_i(t)$, and the orange packet represents a mutated copy of the green packet, hypothetically stored at the node, having accuracy of $0$ (since it is a misleading packet) and version age $X_i(t)$. We are interested in finding the accuracy and version age of the green packets at all nodes. Essentially, now node $i$ sends its green packet to node $j$ with rate $(1-p)\frac{\lambda}{n-1}$ and its orange packet with rate $p\frac{\lambda}{n-1}$. Upon receiving a packet from node $i$, if node $j$ chooses to keep the received packet as per the state reset protocol (thereby forming a new green packet at node $j$), then a new orange packet is also immediately created by mutating the new green packet, such that both the new packets have the same version age/number. 

We now proceed to the SHS characterization, where we select the continuous state as $(\pmb{T}(t),\pmb{X}(t))\in \mathbb{R}^{2n}$, where $\pmb{T}(t)=[T_1(t),\ldots,T_n(t)]$ and $\pmb{X}(t)=[X_1(t),\ldots,X_n(t)]$ denote the instantaneous accuracy and instantaneous version age, respectively, of the green packets stored at the $n$ user nodes at time $t$. Transition $(i,j,h)$ is said to take place when node $i$ sends a packet to node $j$, with $h=0$ depicting that node $i$ sent its green packet (i.e., communicated its packet honestly) and $h=1$ indicating node $i$ sent its orange packet (i.e., communicated a misleading packet), with transition $(0,0,1)$ representing an update at the source. The SHS operates in a single discrete mode with the continuous state obeying the differential equation $(\pmb{\dot T}(t),\pmb{\dot X}(t))=\pmb{0}_{2n}$.  
The set of transitions for this SHS is 
\begin{align}\label{eqn:L-set_of_transitions}
    \mathcal{L}= \{(0,0,1)\} &\cup \{(0,i,1):i \in \mathcal{N}\} \cup \nonumber \\
    &\{(i,j,h):i,j \in \mathcal{N},h \in \{0,1\}\}
\end{align}
such that the transition $(i,j,h)$ resets a state vector $(\pmb{T},\pmb{X})$ at time $t$ to $\phi_{i,j,h}(\pmb{T},\pmb{X},t)\in \mathbb{R}^{2n}$ post transition. The rates $\lambda_{ijh}$ for each transition $(i,j,h)$ are as follows,
\begin{align} \label{eqn:rates_lambda}
\lambda_{ijh} = \begin{cases} 
\lambda_e, & i=0,j=0,h=1\\
\frac{\lambda_s}{n}, & i=0,j\in \mathcal{N},h=1\\
(1-p)\frac{\lambda}{n-1}, & i,j\in \mathcal{N},h=1\\
p\frac{\lambda}{n-1}, & i,j\in \mathcal{N},h=0
\end{cases}
\end{align}

Next, we define some variables that would be useful in our analysis later. Given a set of nodes $A \subseteq \mathcal{N}$ and a continuous state $(\pmb{T},\pmb{X})$, we define $X_A=\min_{j\in A}X_j$, with $X_A=\infty$ if $A=\varnothing$. Let $V(A)=\arg \min_{j \in A} X_j$ denote the set of nodes with lowest version age (note that both the green and orange packets at each node have the same version age). Consider another set of nodes $B$, such that $A \cap B =\varnothing$, i.e., the sets are disjoint. We define $T_{A,B}$ as follows:
\begin{itemize}
    \item If $X_A \leq X_B$ and $1\in \{T_j:j \in V(A)\}$, then $T_{A,B}=1$. For all other cases, $T_{A,B}=0$.
\end{itemize}
Essentially, if we were to collect all the green packets from nodes in set $A$ and all the orange packets from nodes in set $B$, and pick the freshest packets from this collection, such that one of these freshest packets happen to have accuracy of $1$, then $T_{A,B}=1$. Clearly, if $X_A > X_B$, then all the freshest packets are orange packets, which have accuracy of $0$. Further, even if $X_A \leq X_B$, no green packet is guaranteed to have the truth, since the nodes in $V(A)$ could have been misinformed by their previous senders. With these definitions, based on transition $(i,j,h)$ at time $t$, the reset map to $\phi_{i,j,h}(\pmb{T},\pmb{X},t)=[T_1',\ldots,T_n',X_1',\ldots,X_n']\in \mathbb{R}^{2n}$ can be described as
\begin{align}\label{eqn:continuous_state_Tl}
T_{\ell}' = \begin{cases} 
1, & i=0,j\in \mathcal{N},h=1,\ell=j\\
T_{\{i,\ell\},\varnothing}, & i,j\in \mathcal{N},h=1,\ell=j\\
T_{\{\ell\},\{i\}}, & i,j\in \mathcal{N},h=0,\ell=j\\
T_{\ell}, & \text{otherwise}
\end{cases}
\end{align}
and
\begin{align} \label{eqn:continuous_state_Xl}
X_{\ell}' = \begin{cases} 
X_{\ell}+1, & i=0,j=0,h=1,\ell=j\\
0, & i=0,j\in \mathcal{N},h=1,\ell=j\\
\min\{X_i,X_{\ell}\}, & i,j\in \mathcal{N},h=1,\ell=j\\
\min\{X_i,X_{\ell}\}, & i,j\in \mathcal{N},h=0,\ell=j\\
X_{\ell}, & \text{otherwise}
\end{cases}
\end{align}

In the next section, we pick a series of test functions $\psi:\mathbb{R}^{2n}\times [0,\infty) \to \mathbb{R}$ that are time-invariant (consequently we will drop the third input $t$ and write $\psi(\pmb{T},\pmb{X},t)$ as $\psi(\pmb{T},\pmb{X})$), satisfying $\dot \psi(\pmb{T}(t),\pmb{X}(t))=0$, such that their long-term expected value $\mathbb{E}[\psi]=\lim_{t \to \infty} \mathbb{E}[\psi(\pmb{T}(t),\pmb{X}(t),t)]$ will be useful for analysis later. 
Defining $\mathbb{E}[\psi(\phi_{i,j})]=\lim_{t \to \infty} \mathbb{E}[\psi(\phi_{i,j}(\pmb{S}(t),\pmb{X}(t),t))]$, \cite[Thm.~1]{hespanhashs} yields 
\begin{align} \label{eqn:hespanha_eqn}
    0=\sum_{(i,j)\in \mathcal{L}}(\mathbb{E}[\psi(\phi_{i,j})]- \mathbb{E}[\psi] )\lambda_{ij}
\end{align}
where the left side is set to zero due to $\frac{d\mathbb{E}[\psi(\pmb{T}(t),\pmb{X}(t),t)] }{dt} =0$ at large $t$ as the expectation stabilizes. For more details, the reader is encouraged to look at references \cite{Yates21gossip,hespanhashs}.

\section{Misinformation and Version Age Derivations}\label{sec:derivations}

Since the accuracy status evolution process and version age evolution process are statistically identical for all user nodes, in the following analysis, the sets $A_k$ and $B_m$ correspond to arbitrary sets of $k$ and $m$ user nodes with $A_k \cap B_m =\varnothing$. Our first test function is $\psi(\pmb{T},\pmb{X})= T_{A_k,B_m}$, which is modified upon transition $(i,j,h)$ to  $\psi(\phi_{i,j,h}(\pmb{T},\pmb{X},t))=T_{A_k,B_m}'$ and can be characterized using (\ref{eqn:continuous_state_Tl}), (\ref{eqn:continuous_state_Xl}) and (\ref{eqn:hespanha_eqn}) as follows,
\begin{align}\label{eqn:TAkBm}
    &T_{A_k,B_m}'= \nonumber\\
    &\begin{cases}
        1 , & i=0,j\in A_k,h=1\\
        \mathbbm{1}_{ \{T_{A_k,\varnothing}=1,X_{A_k}=0\} }, & i=0,j\in B_m,h=1\\
        T_{A_k,B_{m+1}} & i \in \mathcal{N} \backslash(A_k \cup B_m),j\in A_k \cup B_m,\vspace{-0.1cm}\\&h=0 \\ 
        T_{A_k,B_{m+1}} & i \in \mathcal{N} \backslash(A_k \cup B_m),j\in B_m,h=1 \\
        T_{A_{k+1},B_m} &  i \in \mathcal{N} \backslash(A_k \cup B_m),j\in A_k,h=1 \\
        T_{A_{k+1},B_{m-1}} & i\in B_m,j\in A_k,h=1 \\
        T_{A_k,B_m} & \text{otherwise}
    \end{cases}
\end{align}
In (\ref{eqn:TAkBm}), the effect of transition $i \in \mathcal{N} \backslash(A_k \cup B_m),j\in B_m,h=1$ is interesting. Though node $i$ sends its green packet to node $j$ (as $h=1$ implies honest communication), $T_{A_k,B_m}$ is only concerned with the orange packet at node $j$, which always maintains an accuracy status of $0$. Hence, for the purposes of $T_{A_k,B_m}$, it appears as if node $i$ sent its orange packet to potentially replace the orange packet at node $j$. Since $|\mathcal{N} \backslash(A_k \cup B_m)|=n-k-m$ and $|B_m|=m$, there are $(n-k-m)m$ such unique transitions, each with rate $\frac{\lambda}{n-1}$. Let us define
\begin{align}
t_{k,m}=\lim_{t \to \infty} \mathbb{E}[T_{A_k,B_m}(t)], \ \ c_k= \lim_{t \to \infty} \mathbb{E}[\mathbbm{1}_{ \{T_{A_k,\varnothing}=1,X_{A_k}=0\} }]. \nonumber 
\end{align}
Then, using (\ref{eqn:hespanha_eqn}) with (\ref{eqn:TAkBm}), we obtain
\begin{align}
    0=&(1-t_{k,m})  k\frac{\lambda_s}{n}
    + (c_k -t_{k,m})  m\frac{\lambda_s}{n} \nonumber\\
    &+ (t_{k,m+1} -t_{k,m})  (n-k-m)(pk+m)\frac{\lambda}{n-1} \nonumber\\
    &+ (t_{k+1,m} -t_{k,m})  (1-p)(n-k-m)k\frac{\lambda}{n-1} \nonumber\\
    &+ (t_{k+1,m-1} -t_{k,m}) (1-p)km\frac{\lambda}{n-1}
\end{align}
which upon rearrangement gives
\begin{align}\label{eqn:tkm}
    &t_{k,m}= \nonumber\\
    &\frac{1}{(k+m)\frac{\lambda_s}{n} + (n-k-m)(k+m)\frac{\lambda}{n-1} + (1-p)km\frac{\lambda}{n-1}} \nonumber\\
    &\times \bigg( k\frac{\lambda_s}{n} + c_km\frac{\lambda_s}{n} + t_{k+1,m}(1-p)(n-k-m)k\frac{\lambda}{n-1} \nonumber\\
    &\qquad + t_{k,m+1}(pk+m )(n-k-m)(n-k-m)\frac{\lambda}{n-1} \nonumber\\
    &\qquad + t_{k+1,m-1}(1-p)km\frac{\lambda}{n-1}\bigg)
\end{align}

Note that since $A_k$ and $B_m$ are disjoint sets in (\ref{eqn:TAkBm}), only $t_{k,m}$ satisfying $k\geq 0, m\geq0, k+m\leq n$ are encountered. For example, (\ref{eqn:tkm}) might at first give an impression that $t_{n,0}$ is dependent on $t_{n+1,0}$. However $\mathcal{N} \backslash(A_k \cup B_m)=\varnothing$ for $(k,m)=(n,0)$, hence there are no transitions of type $i \in \mathcal{N} \backslash(A_k \cup B_m),j\in B_m,h=1$ in (\ref{eqn:TAkBm}), making the coefficient of $t_{n+1,0}$ zero in (\ref{eqn:tkm}). Likewise the coefficients of $t_{n,1}$ and $t_{n+1,-1}$ also become zero, giving $t_{n,0}=1$ from (\ref{eqn:tkm}).

\begin{figure}[t]
\centerline{\includegraphics[width=0.4\columnwidth]{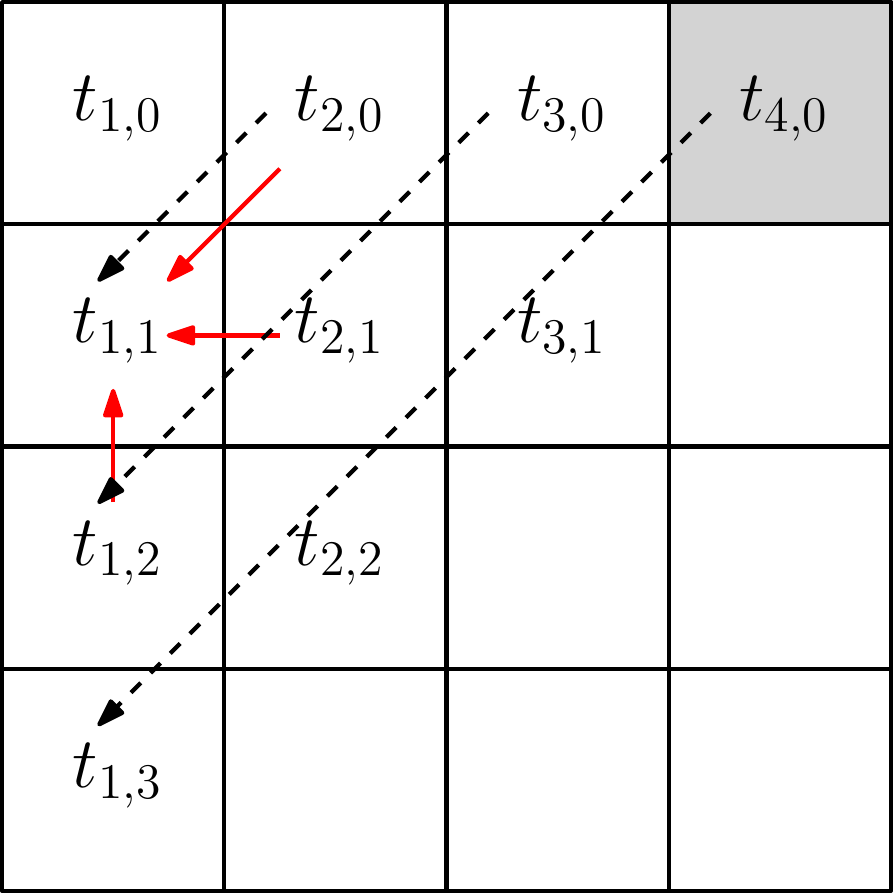}}
\caption{Solving $t_{k,m}$ with dynamic programming for $n=4$. }
\label{fig:dynamic_programming}
\vspace*{-0.7cm}
\end{figure}

Our next test function is $\psi(\pmb{T},\pmb{X})=  \mathbbm{1}_{ \{T_{A_k,\varnothing}=1,X_{A_k}=0\} }$, which has the following $(i,j,h)$ transition map,
\begin{align}\label{eqn:1(TAk=,null)}
    &\mathbbm{1}_{ \{T_{A_k,\varnothing}=1,X_{A_k}=0\} }'= \nonumber \\
    &\begin{cases}
        0 , & i=0,j=0,h=1 \\
        1 , & i=0,j\in A_k,h=1 \\
        \mathbbm{1}_{ \{T_{A_{k+1},\varnothing}=1,X_{A_{k+1}}=0\} }, & i \in \mathcal{N} \backslash A_k ,j\in A_k,h=1 \\
        \mathbbm{1}_{ \{T_{A_k,\varnothing}=1,X_{A_k}=0\} }, & \text{otherwise}
    \end{cases}
\end{align}
Here, if there is a node $j \in A_k$ storing a green packet with accuracy $1$ and version age $0$, i.e., the freshest truth, then $\mathbbm{1}_{ \{T_{A_k,\varnothing}=1,X_{A_k}=0\} }$ will remain $1$ irrespective of any type of transition, since all incoming packets to such node $j$ will be discarded in accordance with the state reset protocol of Section~\ref{sec:sysmodel}. If $\mathbbm{1}_{ \{T_{A_k,\varnothing}=1,X_{A_k}=0\} }=0$, then the only way this test function can change value is if the latest truth is honestly communicated to a node in $A_k$ (i.e., a green packet containing the latest truth is communicated), captured by the second and third cases of the reset map in (\ref{eqn:1(TAk=,null)}).

The corresponding linear equation from (\ref{eqn:1(TAk=,null)}) using (\ref{eqn:hespanha_eqn}) is
\begin{align}
    0=& (0 -c_k)\lambda_e  +  (1 -c_k)k\frac{\lambda_s}{n} \nonumber\\
    & + (c_{k+1} -c_k)  (1-p)k(n-k)\frac{\lambda}{n-1}
\end{align}
which upon rearrangement gives
\begin{align}\label{eqn:ck}
    c_k =\frac{ k\frac{\lambda_s}{n} + c_{k+1}(1-p)k(n-k)\frac{\lambda}{n-1} }{ \lambda_e+ k\frac{\lambda_s}{n}+  (1-p)k(n-k)\frac{\lambda}{n-1}}
\end{align}
Since the laws of accuracy and age processes at all user nodes are the same, from (\ref{eqn:F(t)_def}) we get that the long-term expected fraction of nodes with truth is $F= \lim_{t \to \infty} \mathbb{E}[T_1(t)]=t_{1,0}$. Therefore, for computation of $F$, we first solve for $c_k$ using (\ref{eqn:ck}), starting from $k=n$, with $c_n=\frac{\lambda_s}{\lambda_e + \lambda_s}$ and successively substituting for $k=n-1,\ldots,1$ in an iterative fashion. Once we have computed all $c_k$, we compute $t_{k,m}$ using (\ref{eqn:tkm}) in a dynamic programming fashion, shown in Fig.~\ref{fig:dynamic_programming}. Since in (\ref{eqn:tkm}) $t_{k,m}$ depends on $t_{k+1,m}$, $t_{k,m+1}$ and $t_{k+1,m-1}$, starting with $t_{n,0}=1$, we solve $t_{r-s,s}$ following the order $r=n,\ldots,1$ and $s=0,1,\ldots,r-1$, to reach $t_{1,0}=F$ at the end, as further guided by the arrows along the diagonals in Fig.~(\ref{fig:dynamic_programming}). 

Next, we come to test function $X_{A_k}=\min_{j\in A_k}X_j$, with $v_k=\lim_{t \to \infty}\mathbb{E}[X_{A_k}(t)]$. The $(i,j,h)$ transition map can be written as follows,
\begin{align} \label{eqn:XAk}
X_{A_k}' = \begin{cases} 
X_{A_k}+1, & i=0,j=0,h=1\\
0, & i=0,j\in A_k,h=1\\
X_{A_{k+1}}, & i \in \mathcal{N} \backslash A_k ,j\in A_k,h=1\\
X_{A_{k+1}}, & i \in \mathcal{N} \backslash A_k ,j\in A_k,h=0\\
X_{A_k}, & \text{otherwise}
\end{cases}
\end{align}

Note how information mutation does not impact the version age of a packet, since the transition maps in (\ref{eqn:continuous_state_Xl}) and (\ref{eqn:XAk}) do not depend on $h$. The linear equation for (\ref{eqn:XAk}) using (\ref{eqn:hespanha_eqn}) is
\begin{align}
    0=& (v_k+1-v_k)\lambda_e + (0-v_k)k\frac{\lambda_s}{n} \nonumber\\
    &+ (v_{k+1}-v_k)(1-p)k(n-k)\frac{\lambda}{n-1} \nonumber\\
    &+ (v_{k+1}-v_k)pk(n-k)\frac{\lambda}{n-1}
\end{align}
which upon rearrangement gives
\begin{align}\label{eqn:vk}
    v_k=\frac{\lambda_e + v_{k+1}k(n-k)\frac{\lambda}{n-1}   }{k\frac{\lambda_s}{n}+  k(n-k)\frac{\lambda}{n-1}   }
\end{align}

Defining $x_i= \lim_{t \to \infty} \mathbb{E}[X_i(t)]$, with $x_1=\ldots=x_n$ due to network symmetry, similar to $c_k$, we can backward iterate on (\ref{eqn:vk}) to compute the $x_1=v_1$. 

\section{Analysis and Numerical Results}

\begin{figure}[t]
    \begin{center}
    \subfigure[]{\includegraphics[width=0.49\linewidth]{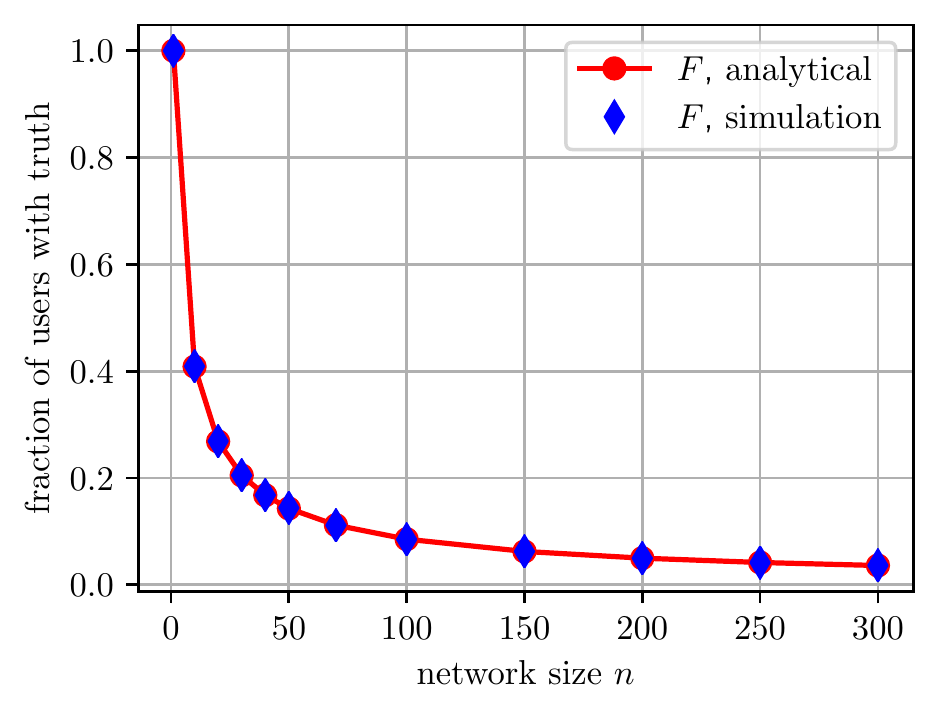}}
    \subfigure[]{\includegraphics[width=0.475\linewidth]{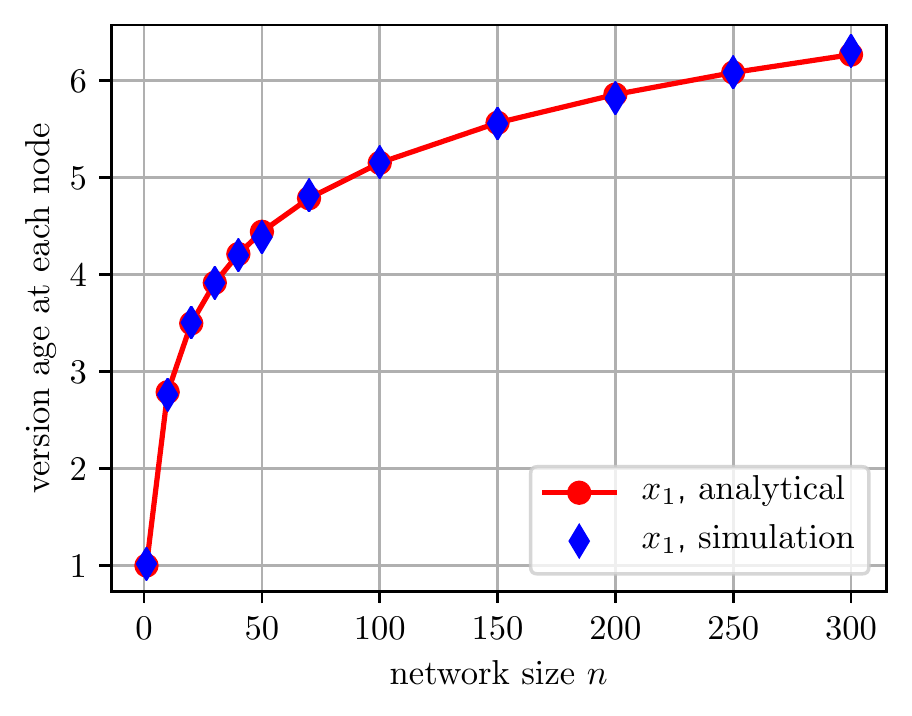}}
    \end{center}
    \vspace{-0.4cm}
    \caption{Analytical (red) and simulation (blue) results compared for $F$ and $x_1$ as a function of $n$, with parameters $\lambda_e=1, \lambda_s= 1, \lambda = 1 ,p=0.9$.}
    \label{fig:simulation_n}
    \vspace{-0.3cm}
\end{figure}

\begin{figure}[t]
    \begin{center}
    \subfigure[]{\includegraphics[width=0.49\linewidth]{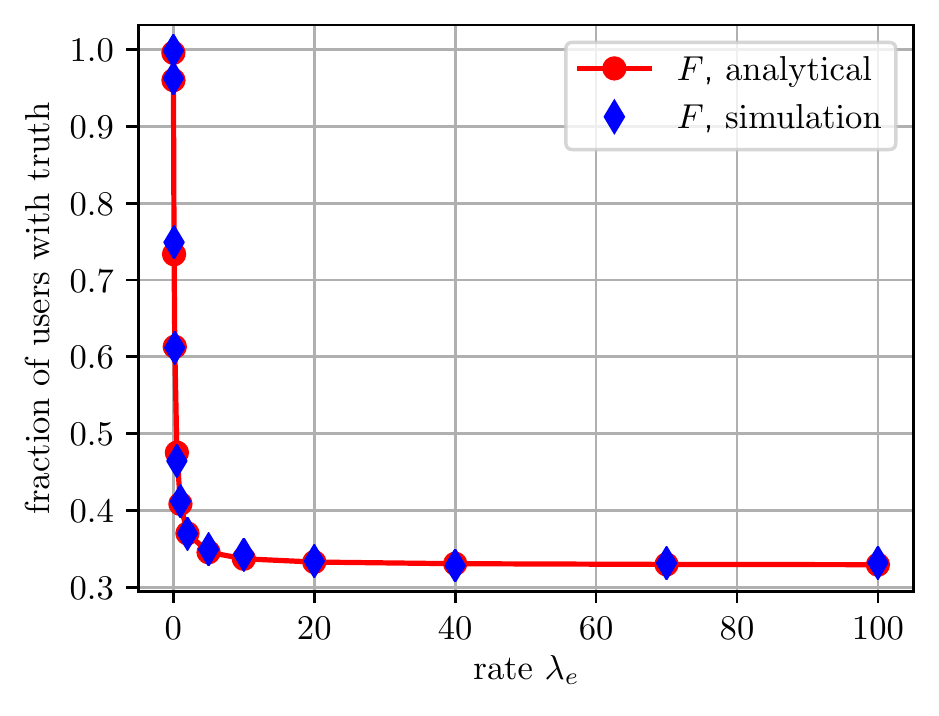}}
    \subfigure[]{\includegraphics[width=0.49\linewidth]{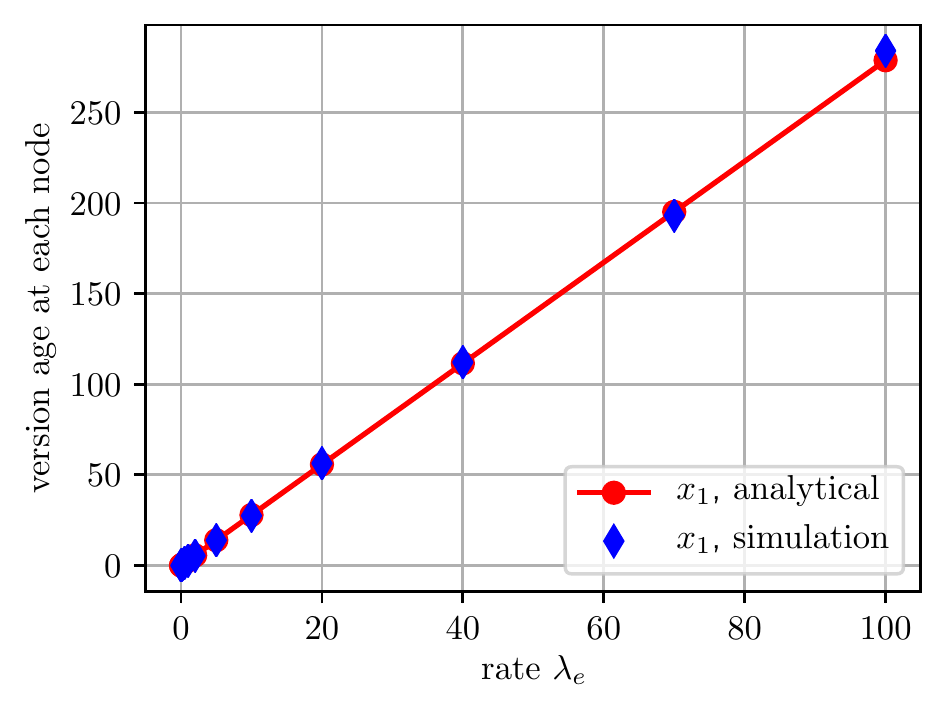}}
    \end{center}
    \vspace{-0.4cm}
    \caption{$F$ and $x_1$ as a function of $\lambda_e$, with $n=10, \lambda_s= 1, \lambda = 1, p=0.9$.}
    \label{fig:simulation_le}
    \vspace{-0.6cm}
\end{figure}

\begin{figure}[t]
    \begin{center}
    \subfigure[]{\includegraphics[width=0.49\linewidth]{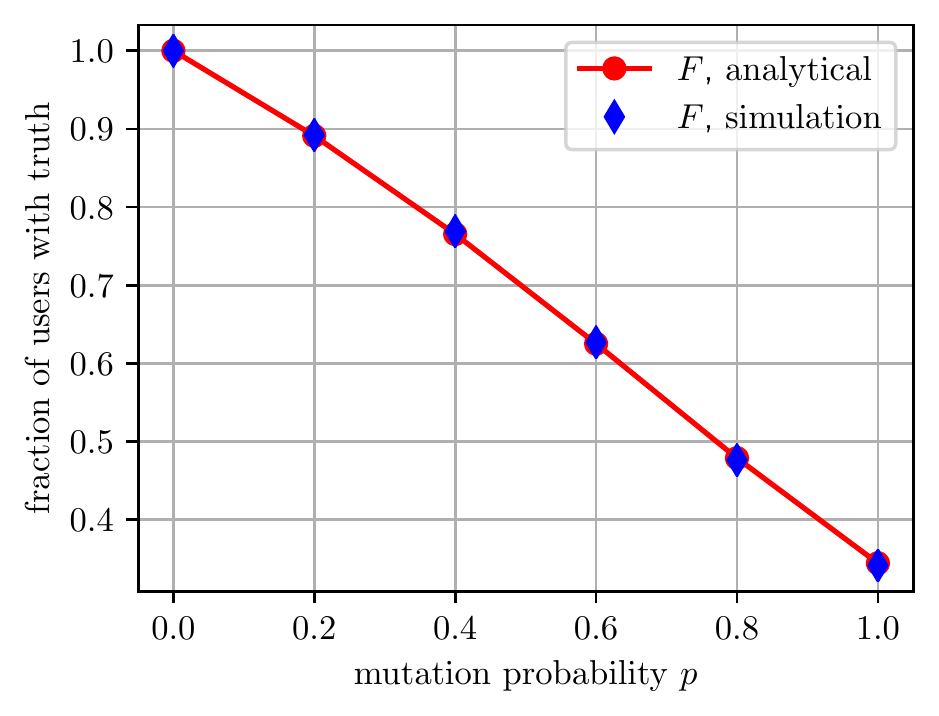}}
    \subfigure[]{\includegraphics[width=0.49\linewidth]{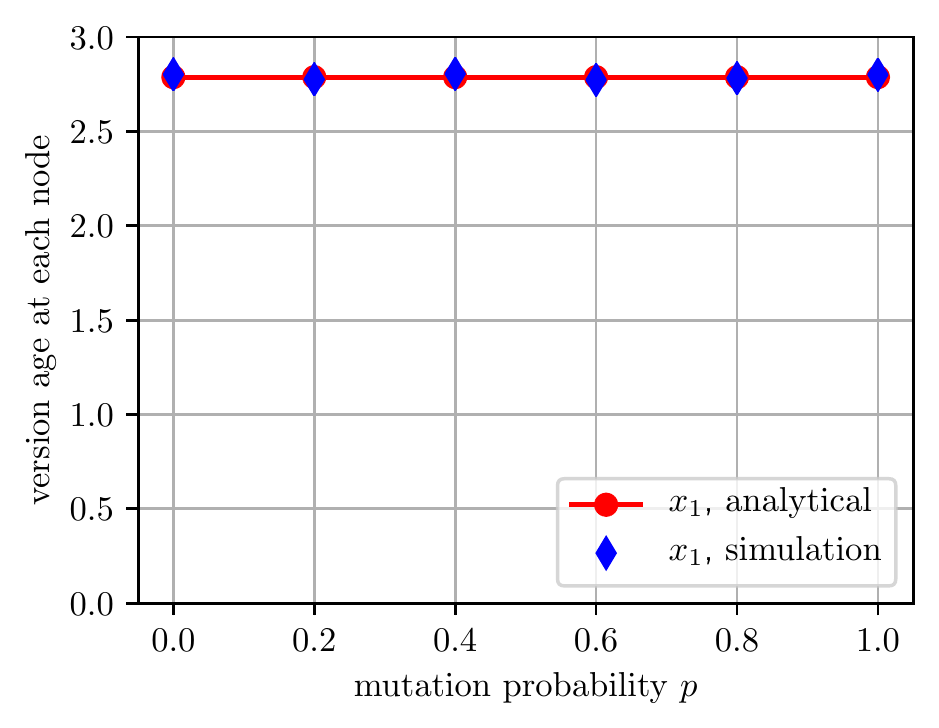}}
    \end{center}
    \vspace{-0.4cm}
    \caption{$F$ and $x_1$ as a function of $p$, with $n=10, \lambda_e=1, \lambda_s= 1, \lambda = 1$.}
    \label{fig:simulation_p}
    \vspace{-0.6cm}
 \end{figure}

\begin{figure}[t]
    \begin{center}
    \subfigure[]{\includegraphics[width=0.49\linewidth]{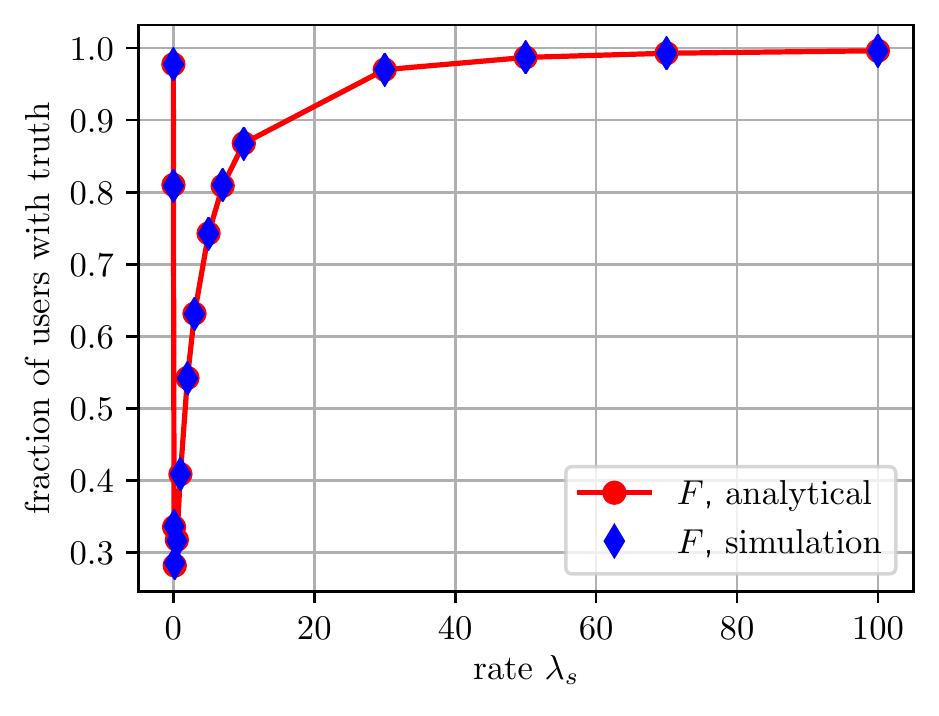}}
    \subfigure[]{\includegraphics[width=0.49\linewidth]{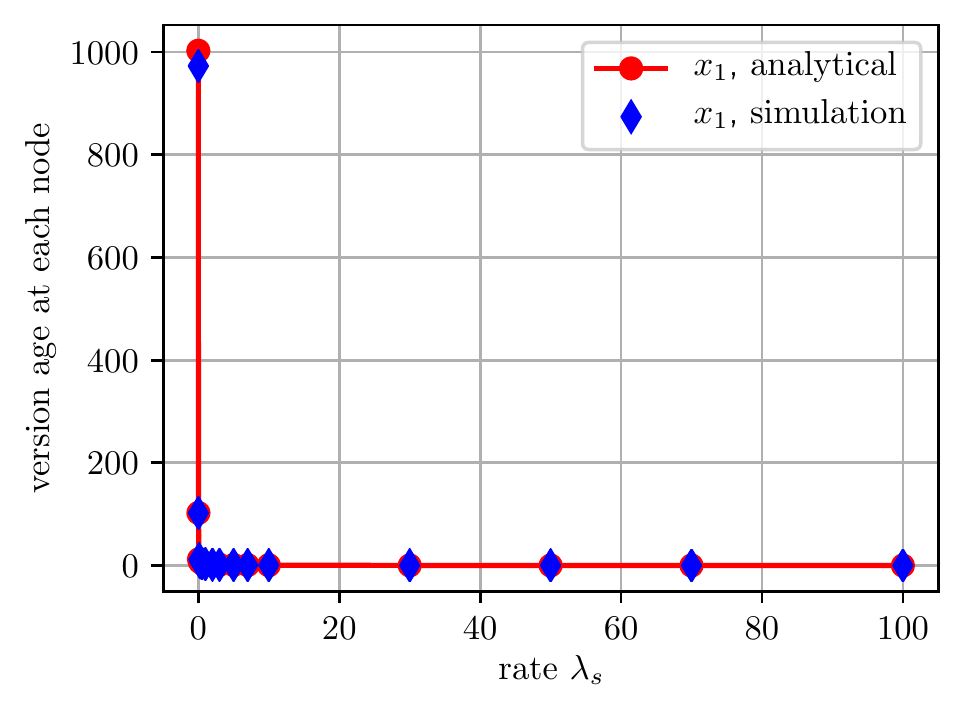}}
    \end{center}
    \vspace{-0.4cm}
    \caption{$F$ and $x_1$ as a function of $\lambda_s$, with $n=10, \lambda_e= 1, \lambda = 1, p=0.9$.}
    \label{fig:simulation_ls}
    \vspace{-0.6cm}
 \end{figure}

\begin{figure}[t]
    \begin{center}
    \subfigure[]{\includegraphics[width=0.49\linewidth]{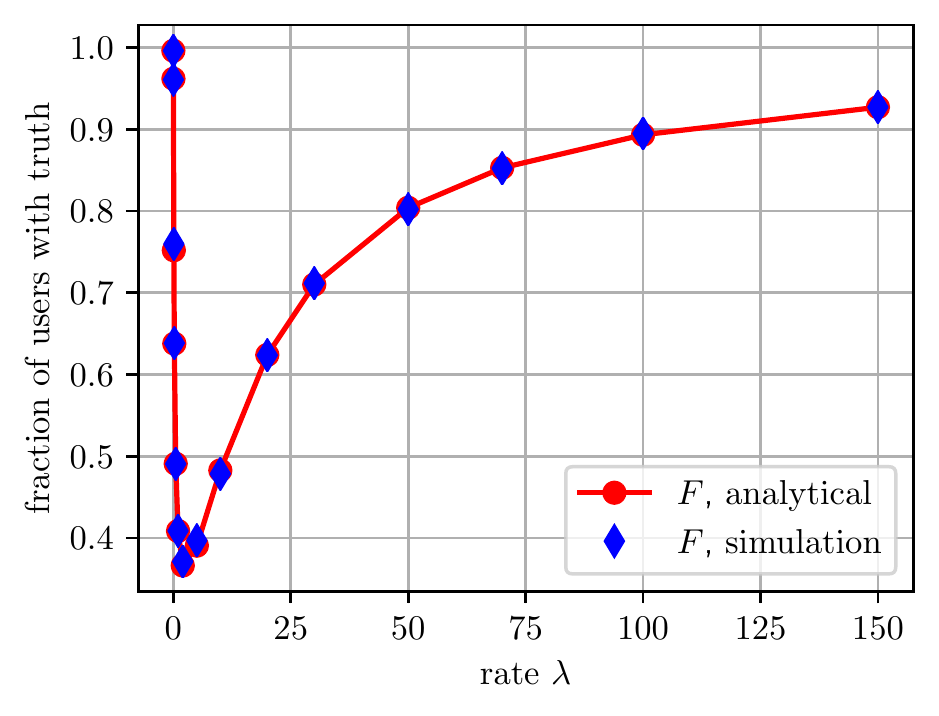}}
    \subfigure[]{\includegraphics[width=0.49\linewidth]{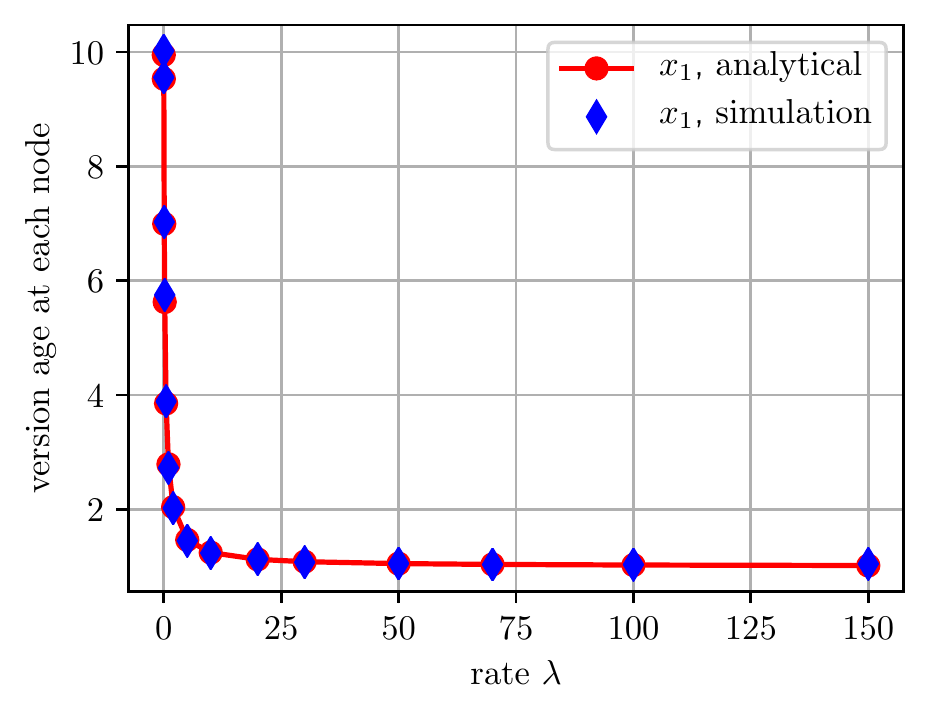}}
    \end{center}
    \vspace{-0.4cm}
    \caption{$F$ and $x_1$ as a function of $\lambda$, with $n=10, \lambda_e= 1, \lambda_s = 1, p=0.9$.}
    \label{fig:simulation_l}
    \vspace{-0.6cm}
\end{figure}

Though it is difficult to derive closed-form expressions for $F$ (expected fraction of nodes with truth) and $x_1$ (expected version age at a node) from the complicated expressions found in (\ref{eqn:tkm}), (\ref{eqn:ck}) and (\ref{eqn:vk}), we provide some interesting observations and further support our theoretical analysis with numerical results. To that end, we simulate the fully connected network model of Fig.~\ref{fig:system_model} for up to a total time of $5\times10^5$ which we use as proxy for $t \to \infty$. We choose parameters $n=10$, $p=0.9$, $\lambda_e=1$, $\lambda_s=1$ and $\lambda=1$, and vary one of the parameters at a time to observe their effects on $F$ and $x_1$, plotting simulation points (blue dots) of $F=t_{1,0}$ on curves (red lines) obtained from equations (\ref{eqn:tkm}), (\ref{eqn:ck}) and (\ref{eqn:vk}) in Figs.~\ref{fig:simulation_n}-\ref{fig:simulation_l}.

Fig.~\ref{fig:simulation_n} shows the plots of $F$ and $x_1$ with respect to the network size $n$. The real-time simulation points coincide with the values derived from the iterative calculations, supporting our theoretical results in Section~\ref{sec:derivations}. Fig.~\ref{fig:simulation_n}(a) shows that $t_{1,0}=1$ when $n=1$, since there is no scope of information mutating in a network containing just one user node due to the absence of inter-node links. Fig.~\ref{fig:simulation_n} suggests that a larger network size leads to more misinformation and staleness at nodes.

Next, Fig.~\ref{fig:simulation_le} shows plots of $F$ and $x_1$ with respect to 
$\lambda_e$. When $\lambda_e$ is small, then substituting $\lambda_e \approx 0$ in (\ref{eqn:ck}) yields $c_k \approx 1$ recursively for $k=n,\ldots,1$. This is because the source hardly gets updated, which causes version age to mostly remain zero at nodes, and due to the prevalence of truth over misinformation, eventually all nodes get the truth corresponding to the version present at the source, i.e., the latest truth. Consequently, by iterating over (\ref{eqn:tkm}), as suggested in Fig.~\ref{fig:dynamic_programming}, we have $t_{k,m}\approx 1$ for all $k,m$ with $k+m\leq n$. 

On the other hand, when $\lambda_e$ is very high, substituting $\lambda_e = \infty$ in (\ref{eqn:ck}) gives $c_k \approx 0$ for all $k$. This is because the current version number at the source is incrementing very fast, causing the versions present at the nodes to become very quickly outdated, driving $x_1$ very large (Fig.~\ref{fig:simulation_le}(b)) and $c_k$ zero. $t_{k,m}$ depends on $\lambda_e$ only through $c_k$, and as large $\lambda_e$ drives $c_k$ to zero, $t_{k,m}$ does become smaller for all $k,m$ (can be seen by iterating in the manner suggested in Fig.~\ref{fig:dynamic_programming}), which suggests that faster version updates at the source causes network nodes to be more misinformed. However, unlike $c_k$, $t_{k,m}$ does not become zero, due to other positive terms in (\ref{eqn:tkm}), owing to the fact that the truth prevails over misinformation, thus many nodes will still have the truth corresponding to older versions.  

Next, if $p$ is very low, then information hardly mutates and only versions of truth circulate in the network. This is supported by substituting $p=0$ in (\ref{eqn:tkm}) where $t_{k,0}$ only depends on $t_{k+1,0}$ giving $t_{k,0}=1$ all for $k=n,\ldots,1$ (see Fig.~\ref{fig:simulation_p}(a)). However when $p$ is high, this does not guarantee that all nodes are misinformed, since the source constantly sends out the latest truth to the network which always gets accepted at the receiving nodes. Substituting $p=1$ in (\ref{eqn:ck}) gives $c_k=\frac{k\frac{\lambda}{n}}{\lambda_s+k\frac{\lambda}{n}}>0$, causing $t_{1,k}$ to have non-zero value for all $k$ from (\ref{eqn:tkm}). From (\ref{eqn:vk}), version age $x_1$ remains independent of $p$, which is also supported by Fig.~\ref{fig:simulation_p}(b).

The dependency of $F$ on $\lambda_s$ and $\lambda$ is more interesting. When $\lambda_s$ is very large, the source updates the network very fast, causing all nodes to have the latest truth at all times. Substituting $\lambda_s=\infty$ in (\ref{eqn:tkm}), (\ref{eqn:ck}) and (\ref{eqn:vk}) gives $c_k\approx 1$, $t_{k,m} \approx 1$ and $x_1\approx 0$ (see Fig.~\ref{fig:simulation_ls}). On the other hand, $v_k$ for all $k$ is a decreasing function of $\lambda_s$ (can be inductively proved from (\ref{eqn:vk}) starting with $k=n$), thus when $\lambda_s$ is very small, by substituting $\lambda_s=0$ in (\ref{eqn:vk}) we get a large version age of $x_1 =\sum_{k=1}^{n-1}\frac{\lambda_e}{k(n-k\frac{\lambda}{n-1})} + \frac{\lambda_e}{\lambda_s} \approx \frac{\lambda_e}{\lambda_s}$ (In Fig.~\ref{fig:simulation_ls}(b), the first simulation corresponds to $\lambda_s=0.001$, for which $x_1=\frac{\lambda_e}{\lambda_s}=\frac{1}{0.001}=1000$). However, interestingly, since the source rarely sends any packet to the network when $\lambda_s$ is close to zero, the whole network continues to gossip about the last version sent by the source to any of the network nodes, and eventually the truth corresponding to that version reaches all nodes. Substituting $\lambda_s=0$ in (\ref{eqn:tkm}) gives  $t_{k,m}=1$, which is also supported by the right extreme of Fig.~\ref{fig:simulation_ls}(a). However, for intermediate values of $\lambda_s$, for  $n\geq 2$, we get $c_k<1$ from (\ref{eqn:ck}), which makes $t_{1,1}<1$, which in turn makes $t_{1,0}$ strictly less than $1$, as also supported by Fig.~\ref{fig:simulation_ls}(a).

Finally, if gossiping rate $\lambda$ is very large, implying that each node has high update sending capacity, then as soon as the source sends an update to a user node, the high gossiping rate allows for instant dissemination of the latest truth to all nodes of the network. This can be verified by substituting $\lambda=\infty$ in (\ref{eqn:tkm}) which gives $t_{k,m}=1$, and can be observed from Fig.~\ref{fig:simulation_l}(a). One would consequently expect the version age to also reduce at all nodes, however, the version age does not become zero in this case, since when the source gets a new version update, the version age at all nodes increments by one, and it continues to have a non-zero value at all nodes until the source sends a packet to some user node in the network. Substituting $\lambda=\infty$ in (\ref{eqn:vk}) gives $x_1=v_1=\ldots=v_n=\frac{\lambda_e}{\lambda_s}$, and in Fig.~\ref{fig:simulation_l}(b), version age converges to $\frac{\lambda_e}{\lambda_s}=1$ as $\lambda$ becomes large.

On the other hand, when $\lambda$ is very small, implying there are negligible inter-node transmissions, then nodes do not receive mutated packets from other nodes and depend primarily on updates received from the source, which always transmits the truth. Substituting $\lambda=0$ in (\ref{eqn:tkm}) gives $t_{1,0}=1$, as also observed in Fig.~\ref{fig:simulation_l}(a). Further, substituting $\lambda=0$ in (\ref{eqn:vk}) gives $x_1=\frac{n\lambda_e}{\lambda_s}$, which evaluates to $10$ for the parameters in Fig.~\ref{fig:simulation_l}(b).

\bibliographystyle{unsrt}
\bibliography{ref_priyanka}

\end{document}